\documentclass[11pt]{article}

\usepackage[final]{acl}

\usepackage{times}
\usepackage{latexsym}
\usepackage{float}
\usepackage[T1]{fontenc}

\usepackage[utf8]{inputenc}
\usepackage{amssymb}

\usepackage{microtype}

\usepackage{inconsolata}

\usepackage{graphicx}
\usepackage{multirow}
\usepackage{booktabs} 
\usepackage{colortbl}
\usepackage{threeparttable}
\usepackage{hyperref}
\usepackage{amsmath}
\title{XLSR-MamBo: Scaling the Hybrid Mamba-Attention Backbone for Audio Deepfake Detection}

\author{
 \textbf{Kwok-Ho Ng\textsuperscript{1}},
 \textbf{Tingting Song\textsuperscript{1}$^{*}$},
 \textbf{Yongdong Wu\textsuperscript{1}},
 \textbf{Zhihua Xia\textsuperscript{1}},
\\
\\
 \textsuperscript{1}College of Cyber Security, Jinan University, Guangzhou, China
\\
 kwokhong@stu2024.jnu.edu.cn, wuyd175@gmail.com, xia\_zhihua@163.com
\\ 
 \small{
   \textbf{$^{*}$Correspondence:} \href{mailto:tingtingsong@jnu.edu.cn}{tingtingsong@jnu.edu.cn}
 }
}

\begin{document}
\maketitle
\begin{abstract}
Advanced speech synthesis technologies have enabled highly realistic speech generation, posing security risks that motivate research into audio deepfake detection (ADD). 
While state space models (SSMs) offer linear complexity, pure causal SSMs architectures often struggle with the content-based retrieval required to capture global frequency-domain artifacts.
To address this, we explore the scaling properties of hybrid architectures by proposing XLSR-MamBo, a modular framework integrating an XLSR front-end with synergistic Mamba-Attention backbones. 
We systematically evaluate four topological designs using advanced SSM variants, Mamba, Mamba2, Hydra, and Gated DeltaNet. 
Experimental results demonstrate that the MamBo-3-Hydra-N3 configuration achieves competitive performance compared to other state-of-the-art systems on the ASVspoof 2021 LA, DF, and In-the-Wild benchmarks. 
This performance benefits from Hydra's native bidirectional modeling, which captures holistic temporal dependencies more efficiently than the heuristic dual-branch strategies employed in prior works.
Furthermore, evaluations on the DFADD dataset demonstrate robust generalization to unseen diffusion- and flow-matching-based synthesis methods. 
Crucially, our analysis reveals that scaling backbone depth effectively mitigates the performance variance and instability observed in shallower models. 
These results demonstrate the hybrid framework's ability to capture artifacts in spoofed speech signals, providing an effective method for ADD. Codes are publicly available at \url{https://github.com/saki-ciallo/XLSR-MamBo}.

\end{abstract}

\section{Introduction}
Driven by rapid advances in generative artificial intelligence (GenAI), cutting-edge text-to-speech (TTS) and voice conversion (VC) technologies have achieved remarkable progress~\citep{casanova2024xtts,du2025cosyvoice,zhou2025indextts2}. The naturalness and timbre similarity of synthesized speech have reached levels that are increasingly challenging for the human ear to distinguish. While beneficial for applications like audiobooks and content creation, these technologies pose severe risks if misused~\citep{li2025survey}. The maturation of zero-shot and few-shot voice cloning allows attackers to synthesize highly realistic forged speech using only short audio segments collected from social media~\citep{azzuni2025voice}. Such misuse not only facilitates the spread of misinformation and erodes societal trust but also threatens voice biometric systems~\citep{barrington2025people, DeepfakesEUNews}. Consequently, developing robust and effective audio deepfake detection (ADD) models is imperative.

Early research in anti-spoofing primarily relied on handcrafted features to detect synthesis artifacts that deviate from human speech production mechanisms~\citep{khan2023battling,yang2025forensic}. 
Although such methods offer interpretability and computational efficiency, these often lack generalization against unseen advanced generators. 
With the advancement of deep learning, mainstream ADD models have shifted to end-to-end architectures, which directly process raw audio for feature extraction and classification.
Regarding front-end feature extraction, XLSR~\citep{babu2022xlsr} is a self-supervised learning (SSL) model based on Wav2Vec 2.0~\citep{baevski2020wav2vec}, pre-trained on massive cross-lingual datasets. 
Benefiting from this large-scale diversity, XLSR has demonstrated competitive generalization capabilities compared to other pre-trained models (PTMs) such as Whisper~\citep{radford2023robust} and WavLM~\citep{chen2022wavlm}, demonstrating strong competitiveness in ADD tasks~\citep{kashyap2025fooling}.
Building on this powerful representation, attention-based classifiers have become an efficient back-end candidate. 
The Conformer architecture~\citep{gulati2020conformer}, which combines the global context modeling of Transformers with the local feature capture of CNNs, is widely adopted. 
By integrating XLSR with Conformer, the XLSR-Conformer~\citep{rosello2023conformer} achieved state-of-the-art (SOTA) performance on the ASVspoof 2021 LA (ASV21LA) and DF (ASV21DF)~\citep{yamagishi2021asvspoof} dataset at the time. 
Subsequent variants, such as XLSR-Conformer+TCM~\citet{truong2024temporal}, XLSR-SLS~\citet{zhang2024audio}, and Kanformer~\citep{dat2025xlsr}, further pushed the performance boundaries by refining the attention mechanism.

However, the quadratic computational complexity of the classical Transformer limits its 
efficiency in modeling dense feature sequences. 
To address this, state space models (SSMs)~\citep{gu2022efficiently} with linear complexity, particularly the Mamba architecture~\citep{gu2023mamba}, have attracted widespread attention. 
RawBMamba~\citep{chen2024rawbmamba} was among the early works to explore the application of Mamba in audio deepfake detection. 
Addressing the unidirectional limitations of the original Mamba, it proposed a Bidirectional Mamba architecture that combines SincLayers/convolutional layers (to capture short-term local features) with bidirectional Mamba blocks (to capture long-range contextual dependencies). 
This design effectively identifies subtle artifacts in synthesized audio and significantly outperforms traditional end-to-end models like Rawformer~\citep{liu2023leveraging}, validating the generalization capability of the Mamba architecture in audio forensics. 
Following this, works like XLSR-Mamba~\citep{xiao2025xlsr} and Fake-Mamba~\citep{xuan2025fake} further advanced the field. The end-to-end BiCrossMamba-ST~\citep{kheir2025bicrossmamba} integrates BiMamba with cross-attention, achieving better performance than RawBMamba.
However, a critical observation in these works is their reliance on manually designed bidirectional strategies (e.g., dual-branch fusion) to approximate the non-causal receptivity of Transformers. 
While these studies successfully validated that substituting Transformer layers with SSM blocks yields efficiency gains, relying solely on this replacement strategy overlooks the unique complementary strengths of different mechanisms.

Although pure SSM models excel in inference efficiency and ADD performance, experimental results from cutting-edge hybrid architectures such as Zamba~\citep{glorioso2024zamba}, Samba~\citep{ren2024samba}, Jamba~\citep{lieber2024jamba}, and Nemotron-H~\citep{blakeman2025nemotron} demonstrate that combining SSMs with Attention yields competitive performance. 
Furthermore, hardware-aware algorithms such as FlashAttention~\citep{dao2022flashattention} and Mesh-Attention~\citep{chen2025mesh} have alleviated the computational bottlenecks of Transformers. 
From the perspective of inductive bias, SSMs specialize in information compression and temporal recurrence, whereas Attention excels at content-based, precise retrieval and associative recall via the induction heads mechanism~\citep{olsson2022context,arora2024simple}. Consequently, in ADD tasks, the Attention mechanism can be better suited to globally correlate specific frequency-domain forgery artifacts.
Specifically, deepfake signals often manifest as a duality: subtle local high-frequency artifacts (requiring the fine-grained temporal recurrence of SSMs) and global spectral inconsistencies (requiring the global retrieval of Attention). 
Relying solely on one mechanism limits the model's ability to capture this full spectrum of manipulation traces~\citep{frank2020leveraging,sun2023ai}.

In light of this, we propose a novel framework, XLSR-MamBo, for anti-spoofing. 
This framework employs XLSR as the front-end and utilizes a hybrid SSM-Attention architecture as the back-end encoder. 
Compared to prior works that rely on manual bidirectionality, we explore advanced variants including Mamba2~\citep{dao2024transformers}, Gated DeltaNet (GDN)~\citep{yang2024gated}, and notably, Hydra~\citep{hwang2024hydra}. 
Hydra introduces a paradigm shift by parameterizing the sequence mixer as a quasiseparable matrix. Unlike heuristic dual-branch extensions, this formulation allows Hydra to perform native bidirectional processing with linear complexity, capturing holistic non-causal dependencies without structural redundancy. This theoretical advantage positions it as an ideal candidate for detecting artifact traces that violate causal consistency.
Furthermore, we investigate the impact of scaling the backbone depth to mitigate inference instability. 
The proposed framework is evaluated on the ASV21LA and DF, In-the-Wild (ITW)~\citep{muller2022does}, and DFADD~\citep{du2024dfadd} datasets, and the results demonstrate its robustness and effectiveness compared to other SOTA systems.

    


\section{Background}

\noindent \textbf{State Space Model.}
The basic SSM defines a continuous-time linear time-invariant (LTI) system. It maps a 1-dimensional input signal $x(t) \in \mathbb{R}$ to a latent state $h(t) \in \mathbb{R}^N$, and then projects it to an output $y(t) \in \mathbb{R}$:
\begin{equation}
    \begin{aligned}
        h^\prime (t) = A h(t) + B x(t), \ 
        y(t) = C h(t)
    \end{aligned}
\end{equation}
where $A \in \mathbb{R}^{N \times N}$ is the state transition matrix, $B \in \mathbb{R}^{N \times 1}$ and $C \in \mathbb{R}^{1 \times N}$ are the input and output projection matrices.

To process discrete sequence data, the continuous system is discretized using a learnable step size $\Delta$, representing the sampling interval. Applying the zero-order hold (ZOH) rule yields the discrete parameters:
\begin{equation}
    \begin{aligned}
        \bar{A} &= \exp(\Delta A) \\
        \bar{B} &= (\Delta A)^{-1}(\exp(\Delta A) - I) \cdot \Delta B
    \end{aligned}
\end{equation}
The discretized SSM state update equation is defined as follows:
\begin{equation}
    \begin{aligned}
        h_t = \bar{A} h_{t-1} + \bar{B} x_t, \ 
        y_t = C h_t
    \end{aligned}
\end{equation}

\noindent \textbf{Selective SSM, Mamba.}
Because the basic LTI SSM uses fixed $\bar{A}$ and $\bar{B}$ matrices across all time steps, it cannot dynamically adjust its memory strategy. Mamba solves this by introducing a selective mechanism where parameters map dynamically from the input, i.e., $x_t \rightarrow (\Delta_t, B_t, C_t)$:
\begin{equation}
    \begin{aligned}
    \Delta_t &= \text{Softplus}(\text{Linear}_{\Delta}(x_t)) \\
    B_t &= \text{Linear}_B(x_t),\ C_t = \text{Linear}_C(x_t) \\
    \bar{A}_t &= \exp(\Delta_t A),\ \bar{B}_t = \Delta_t B_t
    \end{aligned}
\end{equation}
The Mamba state update equation becomes:
\begin{equation}
    \begin{aligned}
        h_t = \bar{A}_t h_{t-1} + \bar{B}_t x_t, \
        y_t = C_t h_t
    \end{aligned}
\end{equation}

\noindent \textbf{State Space Duality (SSD), Mamba2.}
While Mamba's dynamic parameterization enhances expressivity, it reduces computational efficiency. Mamba2 introduces the SSD framework, structurally constraining the continuous state matrix $A$ (e.g., to a scalar multiple of the identity matrix). This simplifies the discretization process and allows the state update to be reformulated as a decay-modulated linear attention, represented by the compact recurrence:
\begin{equation}
    \begin{aligned}
        S_t = \alpha_t S_{t-1} + k_t v_t^\top, \ 
        y_t = q_t^\top S_t 
    \end{aligned}
\end{equation}
where $\alpha_t \in (0,1)$ is a input -dependent decay factor, and $S_t$ is the hidden state matrix at time step $t$. The mapping $x_t \rightarrow (q_t, k_t, v_t)$ generates the query, key, and value vectors corresponding to the attention mechanism. This formulation enables a chunkwise parallel algorithm, significantly accelerating computation via matrix multiplication.

\noindent \textbf{Hydra.}
Standard SSMs are causal recurrent systems. To adapt them for non-causal tasks, Hydra parameterizes the combination of forward and backward scans as a quasiseparable matrix. This mixed matrix contains both lower-triangular (past information propagation) and upper-triangular (future information propagation) semiseparable structures. The Hydra's formula is defined as follow:
\begin{align}
    \text{shift}(SS(X))+\text{flip}(\text{shift}(SS(\text{flip}(X))))+DX
\end{align}
where $X$ is the input sequence, $\text{SS}(\cdot)$ denotes the SSD, $\text{shift}(\cdot)$ and $\text{flip}(\cdot)$ denotes a right-shift and reverse function, and $D = \text{diag}(\delta_1, \dots, \delta_L)$ represents the diagonal elements of the matrix.

\noindent \textbf{Gated DeltaNet.}
To overcome the memory capacity limits of linear attention, GDN integrates the delta rule with Mamba2's dynamic gating decay. This mechanism actively erases components of the prior state that conflict with new information (i.e., those aligned with the current key $k_t$). The GDN recurrence is defined as follows:
\begin{align}
    S_t=\alpha_t(I-\beta_t k_t k_t^\top) S_{t-1} +\beta_t k_t v_t^\top
\end{align}
where $\beta_t$ denotes the input-dependent write strength. The term $S_{t-1} \rightarrow S_{t-1}(I - \beta_t k_t k_t^\top)$ acts as a soft-elimination operator, projecting $S_{t-1}$ into a subspace orthogonal to $k_t$ to minimize interference. For efficiency, the WY representation is used to parallelize the recurrence into chunks, accelerating the training.

\section{Methods}
The original Mamba model performs causal computations in a unidirectional manner, utilizing only historical information. 
However, audio deepfake detection necessitates processing complete utterances to capture global dependencies and spectral inconsistencies. 
While bidirectional adaptations of SSMs can address contextuality, recent advancements in large language models (LLMs) such as Jamba suggest that hybridizing SSMs with Attention mechanisms offers a superior inductive bias compared to pure SSM architectures. 
This hybrid paradigm effectively combines the efficient temporal compression of SSMs with the precise content-based retrieval of Attention.
Yet, the efficacy of such synergistic designs remains underexplored within the specific domain of audio anti-spoofing. 
To systematically assess the potential of this hybrid synergy, we propose a unified modular framework—XLSR-MamBo. 
Unlike previous research that primarily focuses on pure SSM backbones, this work is dedicated to exploring diverse topological combinations of SSM variants within hybrid SSM-Attention architectures.

\begin{figure*}[t]
  \includegraphics[width=\linewidth]{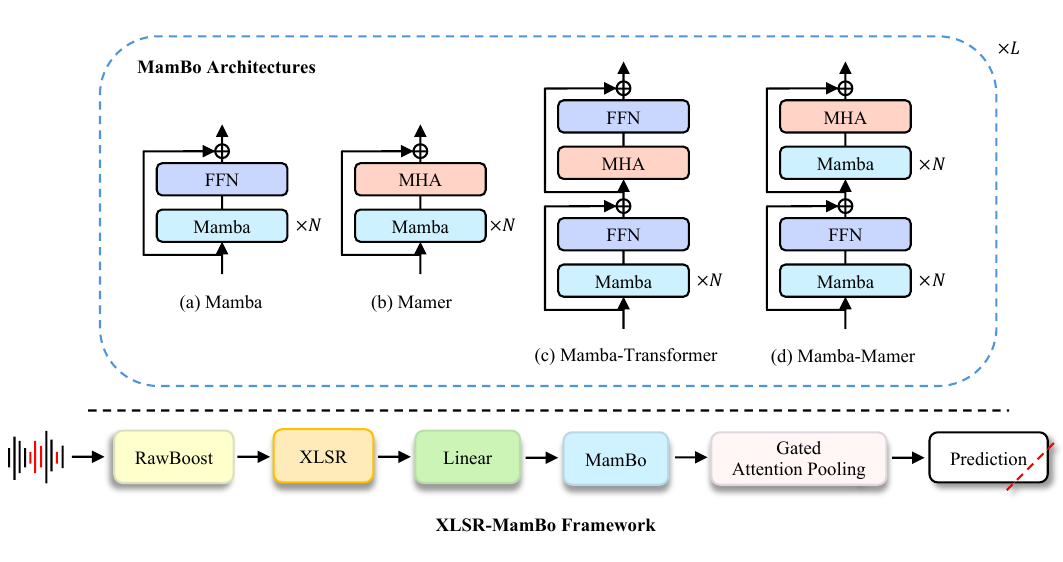}
  \caption{
  The overall proposed XLSR-MamBo framework. The Mamba block is depicted as the representative instantiation of the SSM component. Four variant configurations of the MamBo architectures include: (a) replacing the MHA in Transformer layer with an SSM module, termed Mamba; (b) substituting the FFN in a Mamba layer with MHA to capture non-causal dependencies, termed Mamer; (c) combining Mamba layer with Transformer layers (non-causal and without positional encoding), termed Mamba-Transformer; and (d) hybridizing Mamba layer and Mamer layer, termed Mamba-Mamer.
  }
  \label{fig:overview}
\end{figure*}

The hybrid design is motivated by the complementary strengths of SSMs (efficient sequence modeling) and Attention mechanisms (precise global context recall). Inspired by MambaFormer~\citep{park2024can}, we place the Mamba block after the input projection layer. This configuration leverages the implicit positional encoding inherent in the recurrent structure of SSMs, mitigating the need for explicit positional encodings while maintaining robust sequence modeling~\citep{gu2023mamba}. All blocks incorporate Pre-Norm~\citep{xiong2020layer} connections to enhance training stability.

As shown in Figure~\ref{fig:overview}, we explore four distinct layer architectures. Each structurally resembles a standard Transformer but modifies its internal modules to analyze different syntheses of SSMs and Attention.

\noindent \textbf{MamBo-1} (Mamba Layer, Figure~\ref{fig:overview} (a)): Replaces the multi-head self-attention (MHA) with a Mamba block while retaining the subsequent SwiGLU~\cite{shazeer2020glu} feed-forward network (FFN).
\begin{align}
    H^\prime &= [H +\text{Mamba}(\text{Norm}(H))]^{\times N} \\
    \hat{H} &= H^\prime +\text{FFN}(\text{Norm}(H^\prime))
\end{align}
This provides a baseline to isolate the standalone effectiveness of SSM variants in capturing long-range dependencies without explicit attention.

\noindent \textbf{MamBo-2} (Mamer Layer, Figure~\ref{fig:overview} (b)): Replaces the FFN in the Mamba Layer with a MHA block. This represents a compact intra-block hybrid, placing SSM and Attention in immediate succession. 
\begin{align}
    \hat{H} &= H^\prime +\text{MHA}(\text{Norm}(H^\prime))
\end{align}
This design aims to balance efficient sequence compression with precise global associations within a single block unit.

\noindent \textbf{MamBo-3} (Mamba-Transformer Layer, Figure~\ref{fig:overview} (c)): Adopts a classic interleaved strategy by alternating standard Mamba Layers and Transformer Layers. Building on recent works~\citep{zhang2025mamba} that position Mamba as a promising alternative to self-attention in speech modeling. This sequential hybridization provides a baseline for evaluating the benefits of separating SSM processing and Attention-based refinement into distinct layers.

\noindent \textbf{MamBo-4} (Mamba-Mamer Layer, Figure~\ref{fig:overview} (d)): Further intensifies the hybrid mechanism by alternating Mamba Layers and Mamer Layers. This high-density hybrid explores whether a higher concentration of SSM-Attention interactions yields improved detection of complex spoofing artifacts.

Additionally, we introduce a stacking hyperparameter $N$, inspired by Zamba. This configuration allows for the stacking of $N$ consecutive SSM blocks within a single unit, facilitating a systematic evaluation of depth scaling in SSM modules and its impact on detection performance in audio deepfake detection tasks.

As shown in Figure~\ref{fig:overview}, the overall pipeline of the proposed XLSR-MamBo is structured as follows. The input raw waveform is first processed by the pre-trained XLSR model, which extracts a sequence of high-level speech representations, resulting in a feature matrix $X \in \mathbb{R}^{T \times 1024}$, where $T$ denotes the number of time frames proportional to the input audio duration.
To stabilize training and align feature dimensionality, we apply RMSNorm~\citep{zhang2019root} followed by a linear projection layer, mapping the channel dimension from 1024 to the hidden dimension $D$. The resulting projected features $H \in \mathbb{R}^{T \times D}$ serve as input to the backbone encoder, which consists of stacked MamBo layers ($L$). 
\begin{align}
    H &= \text{Linear}_\text{down}(X)\\
    \hat{H} &= \text{MamBo}(H)^{\times L} \in \mathbb{R}^{T \times D}
\end{align}
This encoder effectively captures complex temporal dependencies and subtle manipulation artifacts while preserving the feature shape $(T \times D)$.
Subsequently, the frame-level features are aggregated into a fixed-dimensional utterance-level representation using gated attention pooling. Finally, a single linear layer projects this pooled representation to a 2-dimensional output, yielding logits for binary classification (spoof vs. bonafide).

\section{Experimental Setup}
\subsection{Datasets and Metrics}
All MamBo variants were trained on the ASVspoof 2019 LA (ASV19LA) training set~\citep{wang2020asvspoof}, which comprises 25,380 training, 24,844 development, and 71,237 evaluation utterances.
To evaluate the generalizability and robustness of the proposed method via cross-dataset evaluation, we conducted experiments on four challenging test sets: ASV21LA (181,566 utterances), ASV21DF (611,829 utterances), ITW (31,778 utterances), and DFADD (1,355 utterances each subset). The ASVspoof 2021 introduces complex transmission channel variations and compression artifacts. The ITW dataset contains real-world deepfake audio sourced from social media, while DFADD includes synthetic speech generated by advanced diffusion- and flow-matching-based TTS models (subsets D1–D3 and F1–F2, respectively).
Following the standard protocols of the ASVspoof Challenges, we report the equal error rate (EER) and the minimum tandem detection cost function (min t-DCF) as evaluation metrics.

\subsection{Implementation Details}
For data augmentation, we used RawBoost~\citep{tak2022rawboost}. Consistent with baseline configurations~\citep{xiao2025xlsr,xuan2025fake}, we trained separate models for the ASV21LA (21LA) and ASV21DF (21DF) evaluations. The 21LA model applied linear and non-linear convolutive noise combined with impulsive signal-dependent noise, while the 21DF model used stationary signal-independent noise.

All audio inputs were processed at 16 kHz with a fixed duration of 4.175s (66,800 samples). We fine-tuned the models using the AdamW~\citep{loshchilov2017decoupled} optimizer ($lr=10^{-5}$, $\beta_1=0.9, \beta_2=0.95$, weight decay=0.05) with a scheduler combining 10\% linear warmup with cosine decay. Training was performed in mixed precision (BF16/FP32) using FocalLoss~\citep{lin2017focal} to mitigate class imbalance, for a maximum of 20 epochs with early stopping (patience of 7 epochs), and a batch size of 32. The top-5 checkpoints with the lowest validation loss were retained. For the architecture, the projection dimension was set to $D=128$, with $L=5$ MamBo layers by default. For the Mamba variants (Mamba2, Hydra, GDN), the state dimension (d\_state) and head dimension were set to 64 and 32, respectively. All experiments were conducted on one RTX 4080 Super 16 GB and two RTX 3090 24 GB GPUs. To facilitate reproducibility, a fixed random seed was used in experimental runs.

\section{Results and Analysis}
\subsection{MamBo-1 and MamBo-2}
\begin{table*}[!htb]
  \centering
  \scriptsize
  \begin{tabular}{lcccccccccc}
    \toprule
    \multirow{2}{*}{\textbf{Model}} & \multirow{2}{*}{$N$} & \multicolumn{2}{c}{\textbf{ASV21LA}} & \textbf{ASV21DF} & \textbf{ITW} & \textbf{D1} & \textbf{D2} & \textbf{D3} & \textbf{F1} & \textbf{F2}\\  
    \cmidrule(lr){3-6} \cmidrule(l){7-11}
    & & min t-DCF $\downarrow$ & EER(\%) $\downarrow$ & EER(\%) $\downarrow$ & EER(\%) $\downarrow$ & EER(\%) $\downarrow$ & EER(\%) $\downarrow$ & EER(\%) $\downarrow$ & EER(\%) $\downarrow$ & EER(\%) $\downarrow$\\  
    \midrule
    \textbf{MamBo-1} \\
    \midrule
    Mamba  & 1 & 0.2178 / {\tiny 0.2196} & 1.19 / {\tiny 1.27} & 2.08 / {\tiny 2.88} & \textbf{4.65} / {\tiny 4.66} & 2.87 / {\tiny 3.28} & 5.17 / {\tiny 5.68}  & 0.15 / {\tiny 0.31} & 11.36 / {\tiny 14.64} & 24.35 / {\tiny 26.47} \\
    ~  & 2 & \textbf{0.2060} / {\tiny 0.2136} & \textbf{0.81} / {\tiny 1.10} & \textbf{2.00} / {\tiny 5.03} & 5.88 / {\tiny 5.96} & 4.14 / {\tiny 4.18} & \textbf{1.18} / {\tiny 1.30}  & 0.51 / {\tiny 0.69} & 30.18 / {\tiny 31.83} & \textbf{20.81} / {\tiny 22.95} \\
    ~  & 3 & 0.2139 / {\tiny 0.2217} & 1.14 / {\tiny 1.43} & 2.15 / {\tiny 2.47} & 5.51 / {\tiny 5.80} & \textbf{1.33} / {\tiny 1.42} & 1.48 / {\tiny 1.48}  & 0.00 / {\tiny 0.19} & \textbf{11.00} / {\tiny 14.56} & 21.69 / {\tiny 22.83} \\
    \cmidrule{2-11}
    Mamba2 & 1 & 0.2170 / {\tiny 0.2209} & 1.21 / {\tiny 1.34} & 2.47 / {\tiny 5.13} & \textbf{4.04} / {\tiny 5.10} & \textbf{0.51} / {\tiny 0.80} & \textbf{0.51} / {1.45}  & 0.00 / {\tiny 0.00} & \textbf{7.16} / {\tiny 18.79} & 16.46 / {\tiny 20.35} \\
    ~  & 2 & \textbf{0.2056} / {\tiny 0.2113} & \textbf{0.79} / {\tiny 0.96} & \textbf{2.01} / {\tiny 2.41} & 5.57 / {\tiny 8.14} & 1.69 / {\tiny 6.43} & 1.69 / {\tiny 4.38}  & 0.00 / {\tiny 0.03} & 9.00 / {\tiny 18.08} & \textbf{12.54} / {\tiny 23.51} \\
    ~  & 3 & 0.2108 / {\tiny 0.2154} & 0.98 / {\tiny 1.13} & 2.37 / {\tiny 3.19} & 5.27 / {\tiny 6.59} & 1.99 / {\tiny 2.34} & 1.84 / {\tiny 2.50}  & 0.00 / {\tiny 0.13} & 7.46 / {\tiny 14.25} & 18.82 / {\tiny 21.88} \\
     \cmidrule{2-11}
    Hydra & 1 & 0.2161 / {\tiny 0.2181} & 1.11 / {\tiny 1.19} & \textbf{2.06} / {\tiny 2.74} & 6.91 / {\tiny 6.99} & \textbf{0.81} / {\tiny 0.95} & \textbf{0.81} / {\tiny 0.98}  & 0.15 / {\tiny 0.15} & 6.20 / {\tiny 7.33} & \textbf{12.48} / {\tiny 13.37} \\
    ~  & 2 & \textbf{0.2100} / {\tiny 0.2140} & \textbf{0.92} / {\tiny 1.06} & 2.10 / {\tiny 3.17} & 7.18 / {\tiny 7.96} & 3.02 / {\tiny 8.61} & 1.03 / {\tiny 5.89}  & 0.00 / {\tiny 2.43} & 16.83 / {\tiny 31.00} & 16.46 / {\tiny 39.47} \\
    ~  & 3 & 0.2109 / {\tiny 0.2300} & 0.97 / {\tiny 1.69} & 2.10 / {\tiny 2.44} & \textbf{5.22} / {\tiny 5.33} & 1.84 / {\tiny 2.03} & 1.69 / {\tiny 2.00}  & 0.00 / {\tiny 0.00} & \textbf{4.20} / {\tiny 6.46} & 13.36 / {\tiny 16.44} \\
     \cmidrule{2-11}
    GDN & 1 & 0.2118 / {\tiny 0.2175} & 1.01 / {\tiny 1.22} & \textbf{2.12} / {\tiny 2.30} & 5.98 / {\tiny 6.08} & 0.81 / {\tiny 0.95} & 0.96 / {\tiny 1.06}  & 0.00 / {\tiny 0.00} & 5.83 / {\tiny 6.65} & 14.53 / {\tiny 15.15} \\
    ~  & 2 & 0.2231 / {\tiny 0.2349} & 1.38 / {\tiny 1.84} & 2.12 / {\tiny 2.71} & 6.29 / {\tiny 6.92} & \textbf{0.66} / {\tiny 0.82} & \textbf{0.96} / {\tiny 1.04}  & 0.37 / {\tiny 0.87} & \textbf{3.17} / {\tiny 6.59} & \textbf{9.30} / {\tiny 12.68} \\
    ~  & 3 & \textbf{0.2110} / {\tiny 0.2140} & \textbf{0.95} / {\tiny 1.08} & 2.68 / {\tiny 3.75} & \textbf{4.16} / {\tiny 4.17} & 3.69 / {\tiny 3.72} & 1.03 / {\tiny 1.15}  & 0.00 / {\tiny 0.00} & 15.65 / {\tiny 16.42} & 20.37 / {\tiny 21.13} \\
    \midrule
    \textbf{MamBo-2} \\
    \midrule
    Mamba  & 1 & 0.2102 / {\tiny 0.2145} & 0.95 / {\tiny 1.11} & 2.54 / {\tiny 2.99} & 6.34 / {\tiny 8.78} & \textbf{1.03} / {\tiny 2.70} & 1.48 / {\tiny 3.38}  & 0.00 / {\tiny 0.09} & 11.15 / {\tiny 16.72} & 21.18 / {\tiny 26.31} \\
    ~  & 2 & \textbf{0.2056} / {\tiny 0.2073} & \textbf{0.79} / {\tiny 0.83} & 2.19 / {\tiny 2.29} & \textbf{4.93} / {\tiny 6.86} & 2.14 / {\tiny 4.09} & 1.33 / {\tiny 3.09}  & 0.00 / {\tiny 0.00} & 7.97 / {\tiny 12.24} & 14.53 / {\tiny 16.78} \\
    ~  & 3 & 0.2138 / {\tiny 0.2172} & 1.05 / {\tiny 1.18} & \textbf{2.09} / {\tiny 2.27} & 6.15 / {\tiny 6.57} & 1.69 / {\tiny 2.09} & \textbf{0.81} / {\tiny 2.21}  & 0.00 / {\tiny 0.00} & \textbf{3.84} / {\tiny 4.40} & \textbf{14.47} / {\tiny 15.83} \\
    \cmidrule{2-11}
    Mamba2 & 1 & 0.2132 / {\tiny 0.2216} & 1.06 / {\tiny 1.37} & \textbf{2.02} / {\tiny 3.35} & 7.23 / {\tiny 7.32} & 1.63 / {\tiny 1.66} & 1.84 / {\tiny 2.11}  & 0.00 / {\tiny 0.00} & 22.14 / {\tiny 24.39} & 23.02 / {\tiny 25.19} \\
    ~  & 2 & \textbf{0.2092} / {\tiny 0.2237} & \textbf{0.90} / {\tiny 1.41} & 2.11 / {\tiny 2.62} & 7.53 / {\tiny 7.62} & \textbf{0.81} / {\tiny 1.04} & \textbf{0.81} / {\tiny 0.91}  & 0.15 / {\tiny 0.15} & 8.34 / {\tiny 8.88} & \textbf{11.36} / {\tiny 12.53} \\
    ~  & 3 & 0.2117 / {\tiny 0.2183} & 1.00 / {\tiny 1.21} & 2.39 / {\tiny 3.14} & \textbf{4.09} / {\tiny 4.12} & 1.48 / {\tiny 1.59} & 1.03 / {\tiny 1.12}  & 0.15 / {\tiny 0.15} & \textbf{6.35} / {\tiny 6.96} & 16.68 / {\tiny 16.97} \\
     \cmidrule{2-11}
    Hydra & 1 & \textbf{0.2066} / {\tiny 0.2199} & \textbf{0.80} / {\tiny 1.32} & \textbf{1.84} / {\tiny 2.73} & 6.24 / {\tiny 6.56} & 1.84 / {\tiny 2.72} & 1.69 / {\tiny 2.37}  & 0.00 / {\tiny 0.03} & \textbf{5.32} / {\tiny 8.54} & 8.85 / {\tiny 23.93} \\
    ~  & 2 & 0.2116 / {\tiny 0.2134} & 1.00 / {\tiny 1.05} & 2.28 / {\tiny 2.94} & 5.01 / {\tiny 5.12} & 1.99 / {\tiny 2.59} & 1.69 / {\tiny 2.56}  & 0.00 / {\tiny 0.00} & 10.33 / {\tiny 13.29} & 20.30 / {\tiny 22.71} \\
    ~  & 3 & 0.2239 / {\tiny 0.2257} & 1.43 / {\tiny 1.48} & 2.08 / {\tiny 2.34} & \textbf{3.80} / {\tiny 5.22} & \textbf{0.37} / {\tiny 1.28} & \textbf{0.37} / {\tiny 1.84}  & 0.00 / {\tiny 0.00} & 5.98 / {\tiny 12.56} & \textbf{6.79} / {\tiny 14.61} \\
     \cmidrule{2-11}
    GDN & 1 & 0.2185 / {\tiny 0.2313} & 1.24 / {\tiny 1.73} & 2.30 / {\tiny 2.38} & 6.52 / {\tiny 6.89} & 1.99 / {\tiny 2.68} & \textbf{1.03} / {\tiny 1.21}  & 0.00 / {\tiny 0.03} & \textbf{7.97} / {\tiny 9.94} & \textbf{17.19} / {\tiny 21.38} \\
    ~  & 2 & \textbf{0.2092} / {\tiny 0.2175} & \textbf{0.89} / {\tiny 1.18} & 1.96 / {\tiny 2.45} & \textbf{5.27} / {\tiny 6.59} & 5.68 / {\tiny 6.44} & 7.97 / {\tiny 8.82}  & 0.00 / {\tiny 0.00} & 24.35 / {\tiny 25.20} & 29.97 / {\tiny 30.83} \\
    ~  & 3 & 0.2124 / {\tiny 0.2156} & 1.01 / {\tiny 1.11} & \textbf{1.94} / {\tiny 2.63} & 7.12 / {\tiny 7.30} & \textbf{1.99} / {\tiny 2.16} & 2.51 / {\tiny 2.66}  & 0.37 / {\tiny 0.39} & 18.16 / {\tiny 19.92} & 27.97 / {\tiny 29.81} \\
    \bottomrule
    \end{tabular}
    
  \caption{
  Comparison of MamBo-1 and MamBo-2 architectures across varying stacking depths $N$ on the ASV21LA, ASV21DF, ITW, and DFADD evaluation sets. All models were trained on ASV19LA. Results are reported as "Best / Avg" across the top-5 checkpoints.
  }
  \label{tab:MamBo-1-2}
\end{table*}

Tables~\ref{tab:MamBo-1-2} and~\ref{tab:MamBo-3-4} summarize the performance of the MamBo series architectures across SSM variants (Mamba, Mamba-2, Hydra, GDN) and SSM stacking depths $N$. To evaluate detection capability against advanced generative algorithms, results on the DFADD dataset are reported separately for the D1-D3 subsets (generated by distinct diffusion models) and F1-F2 subsets (generated by distinct flow-matching models).

In the evaluation of the MamBo-1 architecture on ASV21LA, Mamba, Mamba-2, and Hydra achieved their lowest EERs (0.81\%, 0.79\%, and 0.92\%, respectively) and best min t-DCF scores at a stacking depth of $N=2$. On ASV21DF, however, the variants showed divergent depth preferences: Mamba and Mamba-2 maintained competitive performance at $N=2$, whereas Hydra and GDN demonstrated advantages with a shallower configuration of $N=1$. 
This divergence became more evident on the ITW dataset, where Mamba and Mamba-2 demonstrated stronger generalization at $N=1$, while Hydra and GDN required $N=3$ to achieve the best generalization performance. These data patterns indicate that, within the same architectural framework, different SSM variants exhibit distinct inductive biases, resulting in varying dependencies on stacking depth for generalization to out-of-domain datasets.

Regarding the more challenging DFADD dataset, which presents substantially greater challenges than in-domain datasets, all variants produced comparable and consistently low EERs on D1-D3, demonstrating the model's efficacy in robustly detecting of diffusion-based spoofing attacks. 
On the more difficult F1-F2 subsets, performance of Mamba, Mamba-2, and Hydra varied non-monotonically with stacking depth: strong generalization at $N=1$ and $N=3$, but degradation at $N=2$, suggesting potential optimization challenges at intermediate depths. In contrast, GDN maintained a shallow-layer advantage, achieving its best EERs at $N=1$ (5.83\%) and $N=2$ (3.17\%). The overall data trend suggests that increasing SSM stacking depth generally contributed to improved generalization on complex out-of-domain data, notwithstanding variant-specific fluctuations.

For the MamBo-2 architecture, performance remained relatively balanced across configurations on ASV21LA and DF. On the ITW dataset, however, deeper stacks consistently lowered EERs, with Hydra at $N=3$ attaining the minimum value (3.80\%), representing a relative improvement of approximately 39\% over $N=1$. 
Similar scaling benefits emerged on the challenging F1 subset, where Mamba ($N=3$) and Mamba-2 ($N=3$) achieved substantial EER reductions to 3.84\% (approx. 65\% relative improvement) and corresponding gains of approx. 71\%, respectively. 
The GDN maintained competitive performance in shallower configurations ($N=1$), consistent with observations in MamBo-1. These findings suggest that denser SSM-Attention hybridization amplifies scaling benefits with increased stacking depth, enabling more effective identification of subtle artifacts from unknown generative models.

\subsection{MamBo-3 and MamBo-4}

\begin{table*}[htbp]
  \centering
  \scriptsize
  \begin{tabular}{lcccccccccc}
    \toprule
    \multirow{2}{*}{\textbf{Model}} & \multirow{2}{*}{$N$} & \multicolumn{2}{c}{\textbf{ASV21LA}} & \textbf{ASV21DF} & \textbf{ITW} & \textbf{D1} & \textbf{D2} & \textbf{D3} & \textbf{F1} & \textbf{F2}\\  
    \cmidrule(lr){3-6} \cmidrule(l){7-11}
    & & min t-DCF $\downarrow$ & EER(\%) $\downarrow$ & EER(\%) $\downarrow$ & EER(\%) $\downarrow$ & EER(\%) $\downarrow$ & EER(\%) $\downarrow$ & EER(\%) $\downarrow$ & EER(\%) $\downarrow$ & EER(\%) $\downarrow$\\  
    \midrule
    \multicolumn{11}{l}{\textbf{MamBo-3}} \\
    \midrule
    Mamba  & 1 & 0.2272 / {\tiny 0.2333} & 1.53 / {\tiny 1.77} & 1.80 / {\tiny 2.59} & 7.54 / {\tiny 8.27} & 8.85 / {\tiny 10.56} & 1.69 / {\tiny 2.73}  & 0.00 / {\tiny 0.06} & 16.98 / {\tiny 33.37} & \textbf{18.37} / {\tiny 39.78} \\
    ~  & 2 & 0.2118 / {\tiny 0.2180} & 1.09 / {\tiny 1.30} & \textbf{1.77} / {\tiny 2.68} & \textbf{4.78} / {\tiny 4.82} & \textbf{0.51} / {\tiny 0.63} & \textbf{0.81} / {\tiny 0.91}  & 0.15 / {\tiny 0.18} & \textbf{8.34} / {\tiny 9.03} & 18.67 / {\tiny 20.06} \\
    ~  & 3 & \textbf{0.2107} / {\tiny 0.2149} & \textbf{0.96} / {\tiny 1.14} & 2.01 / {\tiny 2.84} & 5.71 / {\tiny 5.73} & 1.33 / {\tiny 1.45} & 4.80 / {\tiny 5.22}  & 0.15 / {\tiny 0.15} & 9.30 / {\tiny 10.19} & 27.67 / {\tiny 29.59} \\
    \cmidrule{2-11}
    Mamba2 & 1 & 0.2196 / {\tiny 0.2296} & 1.41 / {\tiny 1.64} & \textbf{1.56} / {\tiny 1.88} & 6.04 / {\tiny 6.08} & 1.33 / {\tiny 1.49} & 0.66 / {\tiny 0.82}  & 0.15 / {\tiny 0.25} & 6.20 / {\tiny 6.61} & 12.99 / {\tiny 13.32} \\
    ~  & 2 & 0.2193 / {\tiny 0.2442} & 1.26 / {\tiny 2.20} & 2.15 / {\tiny 3.04} & 4.83 / {\tiny 5.12} & 1.03 / {\tiny 1.30} & 0.66 / {\tiny 1.13}  & 0.00 / {\tiny 0.25} & 7.46 / {\tiny 10.46} & \textbf{10.03} / {\tiny 15.15} \\
    ~  & 3 & \textbf{0.2112} / {\tiny 0.2145} & \textbf{0.94} / {\tiny 1.08} & 2.02 / {\tiny 2.37} & \textbf{4.45} / {\tiny 4.48} & \textbf{0.15} / {\tiny 0.15} & \textbf{0.30} / {\tiny 0.30}  & 0.00 / {\tiny 0.00} & \textbf{3.02} / {\tiny 3.55} & 11.36 / {\tiny 12.50} \\
     \cmidrule{2-11}
    Hydra & 1 & 0.2222 / {\tiny 0.2264} & 1.34 / {\tiny 1.49} & 2.58 / {\tiny 4.79} & 6.00 / {\tiny 6.05} & 2.66 / {\tiny 2.90} & 3.84 / {\tiny 4.04}  & 0.00 / {\tiny 0.00} & 14.47 / {\tiny 15.70} & 17.13 / {\tiny 19.32} \\
    ~  & 2 & 0.2152 / {\tiny 0.2308} & 1.11 / {\tiny 1.68} & 2.12 / {\tiny 2.98} & 5.58 / {\tiny 6.43} & \textbf{0.66} / {\tiny 1.56} & \textbf{1.18} / {\tiny 2.68}  & 0.00 / {\tiny 0.47} & \textbf{3.47} / {\tiny 8.44} & 16.68 / {\tiny 23.57} \\
    ~  & 3 & \textbf{0.2072} / {\tiny 0.2087} & \textbf{0.81} / {\tiny 0.87} & \textbf{1.70} / {\tiny 1.81} & \textbf{4.97} / {\tiny 5.18} & 1.84 / {\tiny 2.53} & 1.33 / {\tiny 1.69}  & 0.00 / {\tiny 0.00} & 11.36 / {\tiny 12.73} & \textbf{16.01} / {\tiny 17.53} \\
     \cmidrule{2-11}
    GDN & 1 & 0.2111 / {\tiny 0.2181} & \textbf{0.96} / {\tiny 1.23} & 2.20 / {\tiny 3.29} & \textbf{4.72} / {\tiny 5.92} & \textbf{1.03} / {\tiny 3.53} & 1.48 / {\tiny 3.30}  & 0.15 / {\tiny 0.18} & 11.51 / {\tiny 17.59} & 12.69 / {\tiny 22.17} \\
    ~  & 2 & \textbf{0.2107} / {\tiny 0.2151} & 0.99 / {\tiny 1.23} & 2.30 / {\tiny 3.15} & 5.54 / {\tiny 6.15} & 4.20 / {\tiny 6.03} & \textbf{1.03} / {\tiny 1.21}  & 0.00 / {\tiny 0.00} & 13.21 / {\tiny 15.18} & \textbf{10.48} / {\tiny 15.19} \\
    ~  & 3 & 0.2127 / {\tiny 0.2152} & 1.00 / {\tiny 1.13} & \textbf{1.82} / {\tiny 1.99} & 5.30 / {\tiny 6.64} & 1.48 / {\tiny 2.16} & 1.03 / {\tiny 2.56}  & 0.30 / {\tiny 0.54} & \textbf{10.18} / {\tiny 18.14} & 15.80 / {\tiny 21.27} \\
    
    \midrule \midrule
    \multicolumn{11}{l}{\textbf{MamBo-4}} \\
    \midrule
    Mamba  &  & 0.2100 / {\tiny 0.2118} & 0.90 / {\tiny 1.00} & 2.14 / {\tiny 2.98} & 5.54 / {\tiny 5.98} & \textbf{1.03} / {\tiny 2.14} & 1.03 / {\tiny 1.18}  & 0.15 / {\tiny 0.29} & \textbf{5.68} / {\tiny 20.85} & \textbf{7.16} / {\tiny 20.72} \\
    Mamba2 &  & 0.2093 / {\tiny 0.2140} & 0.90 / {\tiny 1.06} & 2.17 / {\tiny 2.50} & 6.42 / {\tiny 7.01} & 3.32 / {\tiny 4.65} & 3.69 / {\tiny 5.00}  & 0.15 / {\tiny 0.34} & 16.01 / {\tiny 23.14} & 26.79 / {\tiny 32.85} \\
    Hydra &  & 0.2119 / {\tiny 0.2330} & 0.98 / {\tiny 1.74} & \textbf{1.43} / {\tiny 2.49} & \textbf{5.17} / {\tiny 5.39} & 1.33 / {\tiny 2.70} & 1.84 / {\tiny 4.66}  & 0.00 / {\tiny 0.03} & 14.17 / {\tiny 24.19} & 19.34 / {\tiny 29.30} \\
    GDN &  & \textbf{0.2090} / {\tiny 0.2145} & \textbf{0.86} / {\tiny 1.09} & 2.36 / {\tiny 2.90} & 5.42 / {\tiny 5.72} & 1.69 / {\tiny 1.97} & \textbf{0.66} / {\tiny 1.12}  & 0.00 / {\tiny 0.03} & 7.97 / {\tiny 17.25} & 17.13 / {\tiny 22.59} \\
    \midrule
    \multicolumn{11}{l}{\textbf{MamBo-4} with $L$ = 7} \\
    \midrule
    Mamba  &  & \textbf{0.2084} / {\tiny 0.2162} & 0.93 / {\tiny 1.18} & 2.52 / {\tiny 2.65} & \textbf{4.92} / {\tiny 5.85} & 1.99 / {\tiny 2.70} & 2.81 / {\tiny 3.61}  & 0.15 / {\tiny 0.44} & 14.47 / {\tiny 16.53} & 23.99 / {\tiny 25.69} \\
    Mamba2 &  & 0.2226 / {\tiny 0.2291} & 1.38 / {\tiny 1.62} & 1.94 / {\tiny 2.44} & 5.26 / {\tiny 6.21} & 1.03 / {\tiny 1.42} & \textbf{0.37} / {\tiny 0.51}  & 0.00 / {\tiny 0.23} & 8.49 / {\tiny 9.02} & 11.36 / {\tiny 12.38} \\
    Hydra &  & 0.2090 / {\tiny 0.2123} & \textbf{0.88} / {\tiny 0.99} & \textbf{1.76} / {\tiny 2.45} & 5.60 / {\tiny 5.63} & \textbf{0.51} / {\tiny 0.75} & 1.18 / {\tiny 1.27}  & 0.00 / {\tiny 0.00} & 7.82 / {\tiny 9.79} & 12.69 / {\tiny 15.49} \\
    GDN &  & 0.2142 / {\tiny 0.2244} & 1.11 / {\tiny 1.45} & 1.94 / {\tiny 2.28} & 6.05 / {\tiny 6.29} & 0.66 / {\tiny 1.25} & 0.37 / {\tiny 1.81}  & 0.00 / {\tiny 0.36} & \textbf{5.98} / {\tiny 8.32} & \textbf{9.00} / {\tiny 18.70} \\
    \bottomrule
    \end{tabular}
    
  \caption{
  Comparison of the MamBo-3 architecture across varying stacking depths $N$ and MamBo-4 under different backbone depths ($L=5$ vs. $L=7$, with fixed $N=1$) on the ASV21LA, ASV21DF, ITW, and DFADD evaluation sets. All models were trained on ASV19LA. Results are reported as "Best / Avg" across the top-5 checkpoints.
  }
  \label{tab:MamBo-3-4}
\end{table*}

Table~\ref{tab:MamBo-3-4} further presents results for the more complex hybrid architectures, MamBo-3 and MamBo-4.
On the ASV21LA (MamBo-3), Mamba, Mamba-2, and Hydra attained their lowest EERs (0.96\%, 0.94\%, and 0.81\%, respectively) at $N=3$, with min t-DCF scores generally correlating positively with EER. 
However, GDN optimized EER (0.96\%) at $N=1$ but achieved the best min t-DCF at $N=2$. 
This difference suggests that while the shallower configuration finds a better equilibrium point for the EER, increasing the depth to $N=2$ enhances score calibration, making it more robust under the weighted cost constraints prioritized by the min t-DCF metric.

On the ASV21DF, performance trends diverged across variants: Mamba and Mamba-2 performed best with shallower stacks ($N=2$ for Mamba at 1.77\% EER; $N=1$ for Mamba-2 at 1.56\% EER), effectively addressing compression artifacts. Hydra and GDN, conversely, attained lowest EERs (1.70\% and 1.82\%, respectively) at $N=3$, suggesting that greater depth improves robustness to channel variations.
Generalization on ITW exhibited analogous patterns: Mamba achieved its lowest EER (4.78\%) at $N=2$, Mamba-2 and Hydra at $N=3$ (4.45\% and 4.97\%, respectively).
These results show that increasing model capacity contributes to the effective capture of out-of-domain features.In contrast, GDN maintained its dominance in the shallow layers, achieving a competitive EER of 4.72\% at $N=1$.

\begin{figure}[htbp]
  \includegraphics[width=\columnwidth]{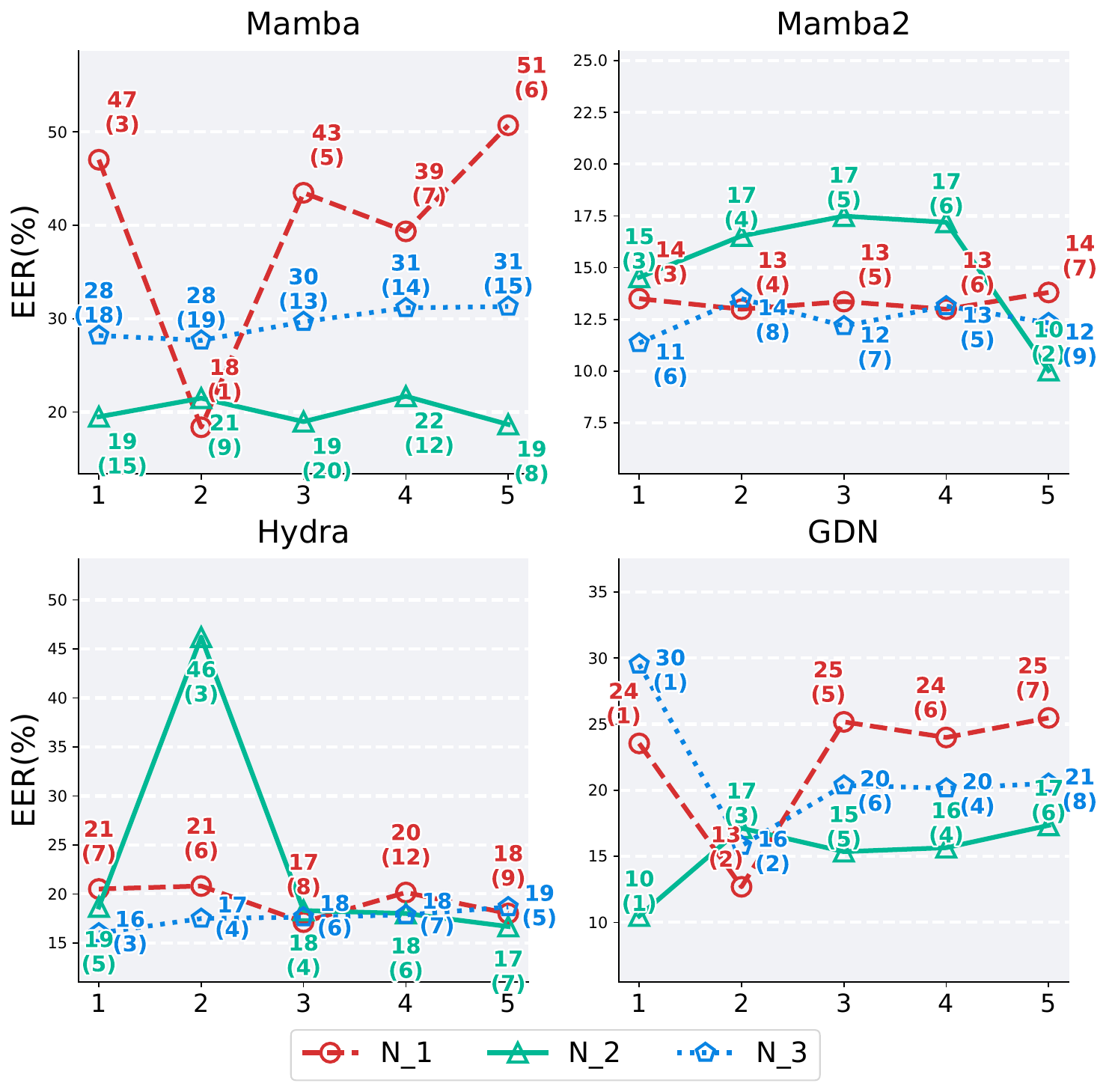}
  \caption{
  Evaluation results of the MamBo-3 architecture utilizing four distinct SSM variants across varying stacking depths $N$ on the DFADD-F2 subset. The x-axis shows the top-5 checkpoints ranked by validation loss (where 1 indicates the lowest loss). Values in parentheses indicate the corresponding training epoch for each checkpoint.
  }
  \label{fig:MamBo-3-line}
\end{figure}

In the detailed evaluation on the DFADD dataset, results on D1-D3 aligned closely with ITW trends. The Mamba ($N=2$) and Mamba-2 ($N=3$) performed best, GDN remained dominated by shallow settings, and Hydra peaked at $N=2$. 
However, on the more challenging F1-F2 subsets, all four variants exhibited improved performance through appropriate stacking compared to shallower settings. The checkpoint EERs trends for the F2 dataset are illustrated in Figure~\ref{fig:MamBo-3-line}. Notably, Mamba-2 achieved the lowest EER of 3.02\% on the F1 dataset with the $N=3$ configuration, while Hydra demonstrated comparable performance of 3.47\% at $N=2$.

\begin{figure}[htbp]
  \includegraphics[width=\columnwidth]{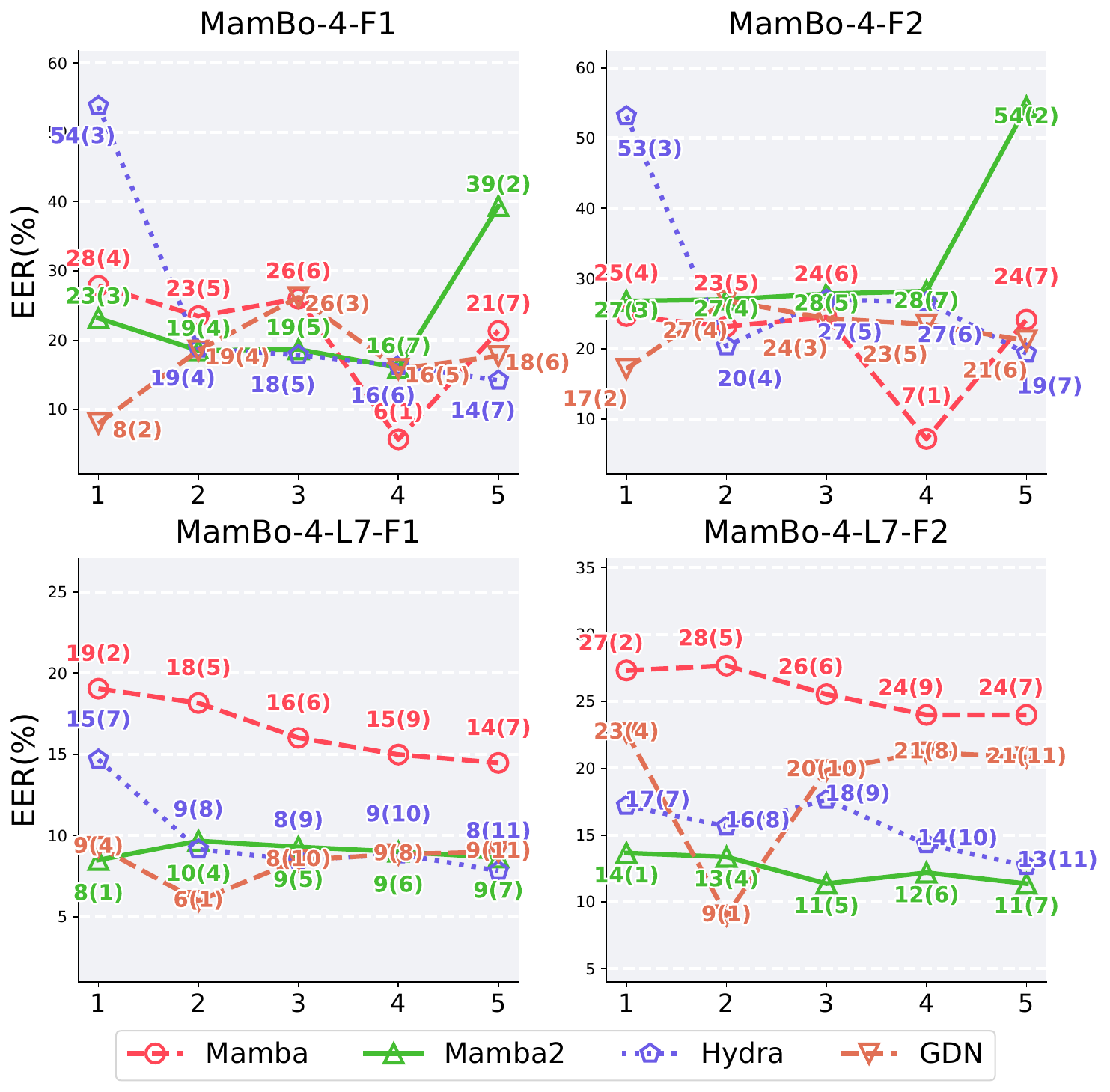}
  \caption{
  Evaluation results of the MamBo-4 architecture integrating four distinct SSM variants on the DFADD-F1 and F2 subsets. The top and bottom rows display performance with backbone depths $L=5$ and $L=7$, respectively (with fixed stacking depth $N$). The x-axis shows the top-5 checkpoints ranked by validation loss (where 1 indicates the lowest loss). Values in parentheses indicate the corresponding training epoch for each checkpoint.
  }
  \label{fig:MamBo-4-line}
\end{figure}

The experimental evaluation of the MamBo-4 architecture involves a comparative analysis between the standard configuration ($L=5$) and a deep ablation setting ($L=7$). Under the default $L=5$ configuration, Mamba and Mamba-2 exhibited comparable performance on ASV21LA, whereas GDN attained the minimum EER of 0.86\%. On the ASV21DF and ITW datasets, Hydra demonstrated notable efficacy, securing the lowest EERs of 1.43\% and 5.17\%, respectively. Regarding the DFADD evaluation, the four SSM variants showed negligible performance differences across the D1-D3 subsets. In contrast, on the F1-F2 subsets, Mamba achieved the lowest EERs (5.68\% on F1, 7.16\% on F2) but exhibited substantial checkpoint instability, with only the best of the top-5 checkpoints achieving single-digit EERs, while the remaining checkpoints showed significant deviation, as shown in Figure~\ref{fig:MamBo-4-line}.

In the ablation with increased backbone depth ($L=7$), deeper architectures generally improved generalization and inference stability across variants on F1-F2 datasets. Notably, Mamba-2 and Hydra demonstrated markedly improved generalization on the F1 dataset, achieving EERs of 8.49\% (approx. 46\% improvement) and 7.82\% (approx. 44\% improvement), respectively. 
However, divergent trade-offs emerged; Mamba performance peaked at $L=5$ and degraded at $L=7$, implying potential overfitting despite improved stability, whereas GDN exhibited enhanced generalization but remained the least stable. 
Overall, Mamba-2 and Hydra demonstrated better cross-domain robustness, with Hydra securing the lowest EERs on ASV21LA (0.88\%) and DF (1.76\%), confirming the efficacy of increasing backbone depth as an optimization strategy.

A comprehensive analysis of the top-5 checkpoint performance reveals persistent low bias accompanied by high variance. 
Specifically, while the top-1 checkpoint achieved minimal validation loss on ASV19LA and improved generalization on DFADD, substantial inference disparity was observed across the remaining candidates. 
This instability indicates that the decision boundaries of shallow architectures remain sensitive to unseen generative algorithms. 
Consistent findings from the MamBo-3 and MamBo-4 experiments on F1-F2 demonstrate that increasing model depth mitigates this issue to varying degrees. 
Consequently, deeper configurations enhance inter-checkpoint consistency and effectively attenuate the performance variance prevalent in shallower models.

\subsection{Compare with Other SOTA Systems}

\begin{table}[ht]
  \centering
  \scriptsize
  \begin{threeparttable}
      \begin{tabular}{lcccc}
        \toprule
        \textbf{Method} & \textbf{Params (M)} & \textbf{21LA} & \textbf{21DF} & \textbf{ITW} \\  
        \cmidrule(lr){3-5}
        & & EER & EER & EER \\  
        \midrule
        RawMamba\tnote{1} & 0.719 & 2.84 & 22.48 & - \\
        RawBMamba\tnote{1} & 0.719 & 3.28 & 15.85 & - \\
        BiCrossMamba-ST\tnote{2} & 0.516 & 3.39 & 14.77 & - \\
        \midrule
        XLSR-Conformer\tnote{3} & 319.74 & 1.38 & 2.27 & -  \\
        XLSR-Conformer+TCM\tnote{4}  & 319.77 & 1.03 & 2.06 & 7.79  \\
        XLSR-SLS\tnote{5} & 341.49 & 2.87 & 1.92 & 7.46 \\
        XLSR-Mamba\tnote{6}  & 319.33 & 0.93 & 1.88 & 6.71  \\
        Fake-Mamba (L)\tnote{7} & 319.72 & 0.97 & 1.74 & 5.85  \\
        \midrule
        \textbf{MamBo-1-Mamba2-N2} & 317.48 & \textbf{0.79} & 2.01 & 5.57 \\
        \textbf{MamBo-2-Hydra-N1} & 316.68 & 0.80 & 1.84 & 6.24 \\
        \rowcolor{gray!20}
        \textbf{MamBo-3-Hydra-N3} & 319.37 & 0.81 & 1.70 & \textbf{4.97} \\
        \textbf{MamBo-4-Hydra-N1} & 318.02 & 0.98 & \textbf{1.43} & 5.17 \\
        \bottomrule
        \end{tabular}
        \begin{tablenotes}
            \scriptsize
            \item[]
            \footnote{1}\citet{chen2024rawbmamba};
            \footnote{2}\citet{kheir2025bicrossmamba};
            \footnote{3}\citet{rosello2023conformer};
            \footnote{4}\citet{truong2024temporal};
            \footnote{5}\citet{zhang2024audio};
            \footnote{6}\citet{xiao2025xlsr};
            \footnote{7}\citet{xuan2025fake}.
        \end{tablenotes}
    \end{threeparttable}
  \caption{
  Comparison of selected configurations from each MamBo variant with SOTA single systems on the 21LA, 21DF, and ITW evaluation sets. Included lightweight end-to-end models that did not use PTM. All baselines trained on ASV19LA.
  }
  \label{tab:base-detile}
\end{table}

Table~\ref{tab:base-detile} presents the performance of the proposed MamBo configurations (with backbone depth fixed at $L=5$) in comparison with state-of-the-art models. 
Experimental data indicates that the MamBo-3-Hydra-N3 hybrid model yields the lowest EERs on the ASV21LA (0.81\%) and ITW (4.97\%) datasets. Regarding aggregate performance, MamBo-3-Hydra-N3 achieves an average EER of 4.75\%, representing a visible improvement over the 5.53\% for the MamBo-4-Hydra-N1 configuration, ref in Table~\ref{tab:MamBo-3-4}.
In comparison to baselines, MamBo-3-Hydra-N3 demonstrates relative improvements of 12.90\% and 9.57\% over XLSR-Mamba on 21LA and 21DF, respectively. Furthermore, it registers a 15.04\% improvement over Fake-Mamba (L) on the ITW dataset. Based on these empirical findings, and considering its parameter efficiency, MamBo-3-Hydra-N3 is selected as the primary backbone for the proposed XLSR-MamBo framework.

\section{Conclusion}
In this paper, we introduced XLSR-MamBo, a modular framework that integrates a pre-trained XLSR front-end with hybrid SSM-Attention back-ends. 
This design relies on complementary inductive biases: leveraging SSMs for efficient compression of local temporal artifacts and Attention mechanisms for the precise retrieval of global spectral inconsistencies. 
Through a systematic evaluation of topological designs and stacking depths, we explored the synergistic potential of this hybridization for ADD.
Experimental results indicate that the MamBo-3-Hydra-N3 configuration achieves competitive performance on the ASV21LA, DF, and ITW datasets, comparable to or surpassing several state-of-the-art single systems with similar parameter counts. 
This performance validates the benefit of Hydra's native bidirectional modeling. 
Theoretically, the quasiseparable matrices employed by Hydra are strictly more expressive than the mixer matrices of heuristic addition-based bidirectional SSMs employed in prior works. 
This enables the model to capture more complex, holistic non-causal dependencies required to distinguish subtle spoofing artifacts without structural redundancy.
Furthermore, evaluations on the DFADD dataset suggest that the framework possesses promising generalization capabilities against advanced diffusion- and flow-matching-based synthesis methods.
Notably, our analysis on scaling properties reveals that increasing backbone depth is essential for mitigating the high performance variance and inference instability often observed in shallower models. Deeper configurations effectively enhance detection robustness across diverse generative algorithms.
Collectively, these findings demonstrate that hybrid SSM-Attention architectures are promising alternatives to traditional Transformer-based or pure SSM-based backbones.
Future work will focus on advancing these hybrid architectures to keep pace with the rapid evolution of speech generation technologies, with the aim of developing more effective and reliable tools for detecting emerging spoofing attacks.

\section*{Limitations}

Despite the promising performance of XLSR-MamBo, this study acknowledges several limitations. 
First, the current evaluation relies exclusively on the single-source training paradigm using the ASV19LA dataset. While this serves as a standard baseline, the absence of large-scale or mixed-source training scenarios limits our assessment of the framework's full scalability. Consequently, it remains unclear whether the reported performance is constrained by data diversity or architectural capacity, and the theoretical upper bound of the model's generalization capability across diverse distributions remains to be fully determined.
Second, the datasets utilized in this work are predominantly English-centric. Although the XLSR front-end benefits from large-scale cross-lingual pre-training, the downstream fine-tuning was restricted to English speech. This potential linguistic bias leaves the model's robustness unverified against deepfakes in languages with distinct phonological structures or tonal characteristics, posing a risk for cross-lingual deployment.
Third, we observed a rapid convergence phenomenon during training. Models frequently achieved minimal validation loss within very few epochs ($< 10$), triggering early stopping. We observed that the checkpoint with the lowest validation loss did not necessarily correspond to the best-performing model for unseen scenarios.
Counterintuitively, these early-stage checkpoints demonstrated competitive generalization capabilities. 
The underlying optimization dynamics governing this relationship between rapid convergence and robust generalization remain an open question and warrant further investigation in future work.

\section*{Ethical Considerations}
While this study utilizes public datasets, the real-world deployment of ADD systems could raise privacy concerns regarding data transmission. Beyond the obvious risks of uploading raw audio, transmitting intermediate representations also presents security vulnerabilities; feature inversion attacks can potentially reconstruct speaker identity and linguistic content from these embeddings. Consequently, future implementations should prioritize privacy-preserving architectures such as on-device processing that fundamentally exclude raw audio transmission and mitigate reconstruction risks.

\section*{Acknowledgments}
This work is supported by the National Natural Science Foundation of China (Grant No. 6227134), Guangzhou Basic and Applied Basic Research Foundation (Grant No. 2025A04J3585). We use AI for writing assistance for paraphrasing and polishing the original content.



\bibliography{custom}

\appendix

\section{Appendix}

\subsection{Datasets}
\label{sec:appendix-datasets}
\begin{itemize}
    \item ASVspoof-2019-LA: \url{https://datashare.ed.ac.uk/handle/10283/3336}
    \item ASVspoof-2021-LA: \url{https://zenodo.org/record/4837263}
    \item ASVspoof-2021-DF: \url{https://zenodo.org/record/4835108}
    \item In-the-Wild: \url{https://deepfake-total.com/in_the_wild}
    \item DFADD: \url{https://huggingface.co/datasets/isjwdu/DFADD}
\end{itemize}

\subsection{Pre-Trained Model}
\label{sec:appendix-ptm}
\begin{itemize}
    \item XLSR-300M: \url{https://docs.pytorch.org/audio/main/generated/torchaudio.pipelines.WAV2VEC2_XLSR_300M.html}
\end{itemize}

\end{document}